\documentclass[twocolumn,aps,prl,showpacs]{revtex4}
\usepackage{graphics}

\begin{document}

\title{Nature of Versatile Chemisorption on TiC(111) and TiN(111) Surfaces}
\author{Carlo Ruberto}
\email{ruberto@fy.chalmers.se}
\author{Aleksandra Vojvodic}
\author{Bengt I. Lundqvist}

\affiliation{Department of Applied Physics, Chalmers University of Technology, 
SE-412 96 G\"oteborg, Sweden}

\begin{abstract}

Density-functional calculations on the polar TiX(111) (X = C, N) 
surfaces show (i) for clean surfaces, strong Ti$3d$-derived 
surface resonances (SR's) at the Fermi level and X$2p$-derived 
SR's deep in the upper valence band and (ii) for adatoms in periods 
1--3, pyramidic trends in atomic adsorption energies, peaking at 
oxygen (9 eV). A concerted-coupling model, where adatom states couple 
to both kinds of SR's in a concerted way, describes the adsorption.  
The chemisorption versatility and the general nature of the model 
indicate ramifications and predictive abilities in, {\it e.g.}, 
growth and catalysis.  

\end{abstract}

\date{10 Apr.\ 2006. Submitted to Phys.\ Rev.\ Lett.
[Applied Physics Report 2005-12 MST]}


\maketitle


In surface processes, like catalysis and film growth, and in materials 
design [through, {\it e.g.}, chemical- (CVD) and physical-vapor deposition 
(PVD), or molecular-beam epitaxy (MBE)], initial adsorption 
is of key importance.  Transition-metal carbides and nitrides (TMX; X = C, N) 
\cite{Oyama} are widely used as substrates, coatings, and interlayers 
in, {\it e.g.}, cutting tools, electronics, catalysis, and biomaterials, 
and as components in the ``MAX'' phases \cite{Barsoum}.  
A study of the nature of adsorption on TMX's is thus highly motivated.  
Generally, TiC and TiN are considered to be models for these materials. 

This Letter reports on an extraordinarily rich adsorptive variation 
on TiX(111) surfaces, found within a systematic 
density-functional-theory (DFT) study of CVD Al$_2$O$_3$/TiX 
coatings on cutting tools \cite{kappaAl2O3}. Our trend calculations on 
adatoms in periods 1--3 of the table of elements show 
large variety in chemisorption strength and diffusion barriers 
(Fig.\ 1).  
To understand this, a detailed analysis of calculated 
bulk, surface, and adsorbate electronic structures is performed. 
The adsorption-energy ($E_{\rm ads}$) trends are correlated with 
electron-structure trends:  
adsorbate-induced differences in surface densities of states 
[$\Delta$DOS($E$)], real-space visualizations of state-resolved DOS's 
[DOS(${\bf r}, E$)], and valence-charge distributions, 
yielding a model for the nature of chemisorption on TiX(111).  
The method relies on DFT calculations performed with the 
planewave-pseudopotential code {\tt dacapo} \cite{dacapo}, with 
PW91 GGA exchange-correlation functional.  
Details are given in Ref. \onlinecite{Ruberto_Vojvodic_Lundqvist}.

\begin{figure}
\scalebox{.4}{\includegraphics{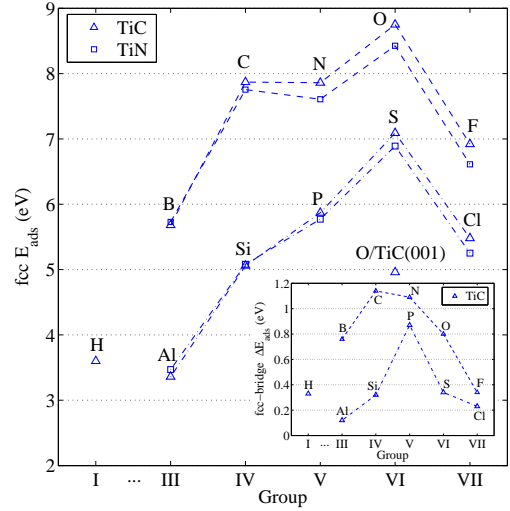}}
\caption{\label{fig:Eads}Calculated atomic adsorption energies $E_{\rm ads}$ 
for fcc site on TiX(111) and for O in top-C site on TiC(001).  
Inset: energy differences between fcc and bridge sites on TiC(111).}
\end{figure}

Our analysis explains chemisorption on TiX(111) as arising from the 
concerted coupling between adatom states and different types of 
TiX surface resonances (SR's) (Fig.\ 2): 
(i) a Ti-localized Fermi-level ($E_F$) SR (TiSR), 
which provides strong chemisorption through overlap 
with adatom frontier orbitals, giving well separated 
bonding (below the adatom level) and antibonding levels (above the 
TiSR and thus $E_F$) and (ii) X-localized SR's 
(XSR's) in the lower part of the surface upper valence band (UVB), 
which provide additional chemisorption, with strength depending on the 
energy separation between XSR's and TiSR-modified adlevel. 
This concerted-coupling model (CCM) shows a decisive but earlier neglected 
role of subsurface X atoms for chemisorption. 
Its generality should make it relevant for other, similar, substrates.

\begin{figure*}
\scalebox{.65}{\includegraphics{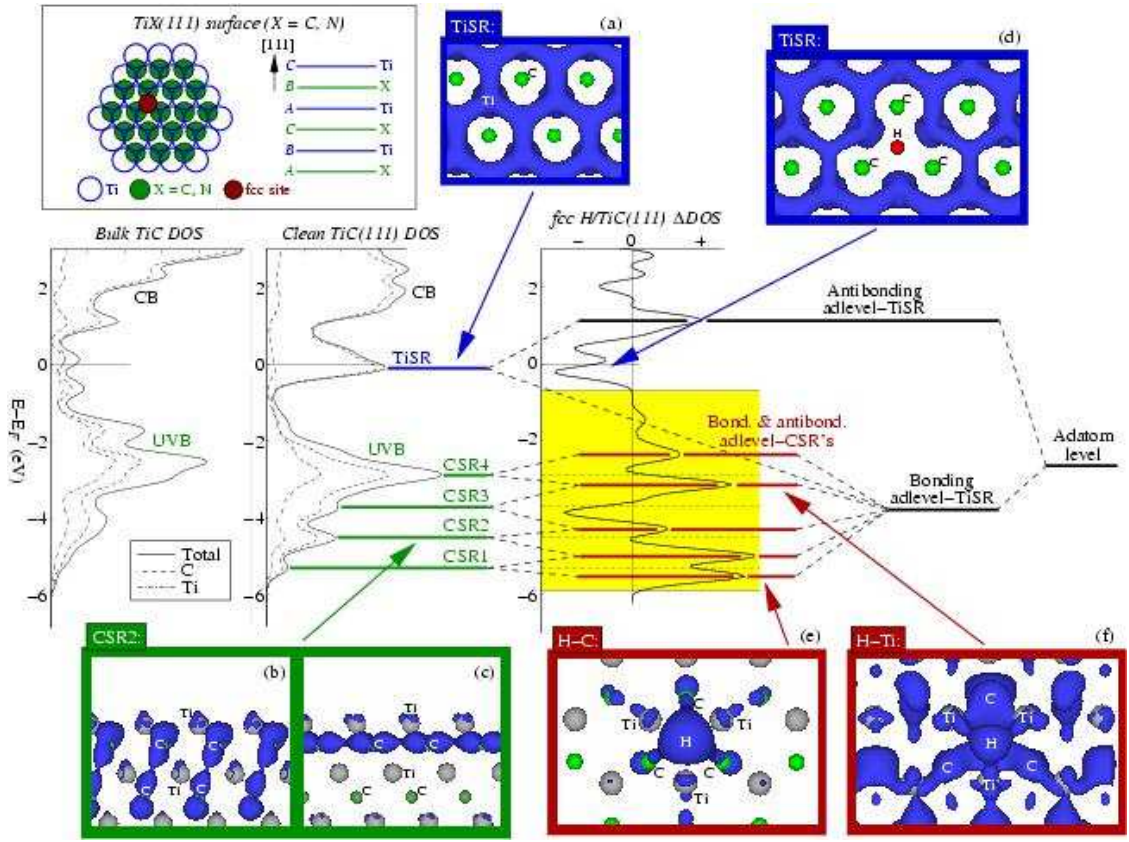}}
\caption{\label{fig:CCM}(Color online).  
Schematic diagram of the concerted-coupling model (CCM), 
as illustrated by fcc H on TiC(111).  Included are: bulk-TiC DOS($E$), 
clean-surface structure and DOS($E$), H/TiC $\Delta$DOS($E$) ({\it cf.}\ 
Fig.\ 3), and DOS(${\bf r}, E$) plots for (a--c) clean 
TiC(111) and (d--f) H/TiC(111), corresponding to the states indicated by 
the arrows [(b--c) are viewed perpendicular to the surface, (a) and 
(d--f) from above the surface, showing only the surface TiC bilayer].}
\end{figure*}

The NaCl-structured TiX's are covalent, ionic, and metallic 
\cite{Schwarz}.  The clean, polar, TiC(111) surface, that is, prepared by 
flash heating under ultra-high vacuum (UHV) conditions, is 
Ti terminated (Fig.\ 2, inset) \cite{Aono,Zaima,Tan} and 
chemically very active, with dissociative adsorption of H$_2$, O$_2$, and 
methanol and a sticking for oxygen 100 times stronger than on the stable, 
non-polar, and less reactive TiC(001) 
\cite{Zaima,Bradshaw,Edamoto}. 
The TiC(111) polarity is stabilized by a compensating 
surface charge provided by the partial filling of a strong TiSR at 
$E_F$ \cite{Aono,Zaima,Bradshaw,Tsukada}. 
For TiN(111), experiments are scarce, while theory indicates 
an environment-dependent termination 
\cite{Ruberto_Vojvodic_Lundqvist,Gall}.  To allow comparisons, Ti 
termination is here used for both (111) surfaces. 

For bulk TiX, our results confirm and extend earlier findings 
\cite{Schwarz}.  Strength, structure, and hardness result from 
directed iono-covalent Ti$3d$--X$2p$ bonds, with energies in an 
X-dominated UVB (bulk TiC DOS in Fig.\ 2). 
Corresponding antibonding Ti$3d$--X$2p$ states lie in a 
Ti-dominated conduction band (CB), separated from the UVB by a pseudogap, a 
region of low but nonzero density of nonbonding Ti$3d$ states.  In TiC, 
$E_F$ falls in the middle of the pseudogap (Fig.\ 2). 
In TiN, the extra electron per formula unit populates the CB, 
pushing UVB and CB to lower energies 
\cite{Schwarz,Ruberto_Vojvodic_Lundqvist} 
and lowering the cohesion energy [14.8 (13.8) eV for TiC (TiN)]. 
Higher electronegativity of N makes TiN more ionic, 
yielding a higher X-localization of the TiN UVB 
\cite{Schwarz,Ruberto_Vojvodic_Lundqvist} 
and a higher Ti$\rightarrow$X Bader \cite{Bader} 
charge transfer [1.5 (1.6) electrons for TiC (TiN)]. 
Our DOS(${\bf r}, E$)'s show that the bonding Ti$3d$--X$2p$ states 
lie mainly in the middle-upper UVB region, while the lower UVB region 
is dominated by bonding X$2p$--X$2p$ states 
\cite{Ruberto_Vojvodic_Lundqvist,Matar}.  
Nonbonding X$2s$ states form a separate lower valence band 
(LVB) at $-9.5$ eV. 

For TiX(111), our DOS($E$)'s (TiC in Fig.\ 2, TiN in Ref.\ 
\onlinecite{Ruberto_Vojvodic_Lundqvist}) show significant Ti$3d$ SR's 
around $E_F$, more filled on TiN due to its higher ionicity.  
DOS(${\bf r}, E$)'s [Fig.\ 2(a)] show that the TiSR's are 
strongly localized around the surface Ti atoms and extend toward the 
fcc sites, {\it i.e.}, what would be X sites in the bulk, 
there coupling with other surface Ti atoms.  
Thus, they are dangling bonds resulting from 
the cleavage of iono-covalent Ti--X states, which downshifts Ti-localized 
CB states and quenches X-localized states in the upper UVB region 
(Fig.\ 2). 
Cleavage modifies also the lower half of the UVB.  The surface DOS($E$)'s 
have three peaks below the main UVB peak, as opposed to one in the bulk 
(Fig.\ 2). DOS(${\bf r}, E$)'s show 
here a mixture of X--X bonding states, strongly localized to the surface 
but also extending into the bulk 
[Figs.\ 2(b--c) illustrate this, with similar plots also 
for the other UVB peaks \cite{Ruberto_Vojvodic_Lundqvist}]. 
Thus, the lower UVB region contains XSR's, causing peak-structure changes 
in the surface DOS($E$)'s (XSR1--XSR4, see Fig.\ 2). 

Our adsorption calculations include full relaxation of all considered 
adatoms in fcc, hcp, and top sites on TiX(111) and of oxygen in top-Ti 
and top-C sites on TiC(001).  The calculated $E_{\rm ads}$ 
values \footnote{The reference points are given by the energies of the 
isolated spin-polarized atoms in their multiplet ground state.} show 
the following trends (Fig.\ 1) 
\cite{Ruberto_Vojvodic_Lundqvist}: (i) much weaker O chemisorption 
on TiC(001) than on TiC(111); (ii) for all adatoms on TiX(111), 
preference for fcc site, followed by hcp; 
(iii) along both periods, pyramidic $E_{\rm ads}$ trends, 
peaking at group-VI elements (O and S); (iv) within each group, 
stronger bonds in period 2 than in period 3; 
(v) very close C and N values; (vi) very similar TiC and TiN values; 
(vii) a striking crossing of the TiC and TiN trends; and 
(viii) large variations in 
estimated fcc--hcp diffusion barriers ({\it i.e.}, $E_{\rm ads}$ 
differences between fully-relaxed fcc/hcp sites and perpendicularly-relaxed 
bridge sites \footnote{Free relaxation brings all bridge adatoms to the fcc 
site.}).  

A combined analysis of calculated $\Delta$DOS($E$)'s and 
DOS(${\bf r}, E$)'s reveals the nature of chemisorption. 
Key features are illustrated by the results for fcc H on TiC(111), showing 
(Fig.\ 2): (i) a strong negative Ti$d$-derived double peak 
around $E_F$, corresponding to a TiSR depletion around the H adatom 
[Fig.\ 2(d)]; (ii) a strong positive Ti$d$-derived peak at 
$+1.1$ eV; (iii) a strong negative C$p$-derived peak close to the 
substrate CSR3 peak ($-3.9$ eV); (iv) a 
broad H$s$-derived positive region between $-6.1$ and $-4.1$ eV 
(peaks at $-5.5$, $-5.0$, and $-4.3$ eV), 
characterized by strong H--C bonding states [Fig.\ 2(e)] 
and weaker H--Ti contributions; and (v) a 
positive region between $-3.6$ and $-0.7$ eV (peaks at $-3.1$, $-2.3$, 
and $-1.4$ eV), 
characterized only by H--Ti bonding states [Fig.\ 2(f)]. 

This evidences a strong adatom--TiSR coupling that depletes 
the TiSR at $E_F$ and yields antibonding states above $E_F$ (at $+1.1$ eV) 
and bonding states below (positive $\Delta$DOS regions between $-6.1$ and 
$-0.7$ eV).  Also, couplings to CSR's are revealed, since 
(i) the strong broadening and splitting of the bonding adatom--TiSR states, 
(ii) the presence of negative peaks in the C-projected $\Delta$DOS at the 
substrate CSR peak energies, and (iii) the presence of bonding adatom--C 
states in only the lower UVB region, successively stronger for 
lower-energy peaks, indicate that the positive $\Delta$DOS peaks 
in-between the substrate UVB peaks correspond to combinations of bonding and 
antibonding adatom--CSR states (successively more bonding for 
lower-energy peaks).  
Thus, the adatom couples in a concerted way to both TiSR and CSR's. 
This CCM is summed up in Fig.\ 2.  

Similar analyses for the other adatoms on TiC and TiN 
(Fig.\ 3 and Ref.\ 
\onlinecite{Ruberto_Vojvodic_Lundqvist}) corroborate the CCM. 
However, changes in adatom group number (III$\rightarrow$VII), 
period number (2$\rightarrow$3), and substrate (TiC$\rightarrow$TiN) 
affect the calculated electronic structures for adatoms in periods 2--3:

\begin{figure}
\scalebox{.61}{\includegraphics{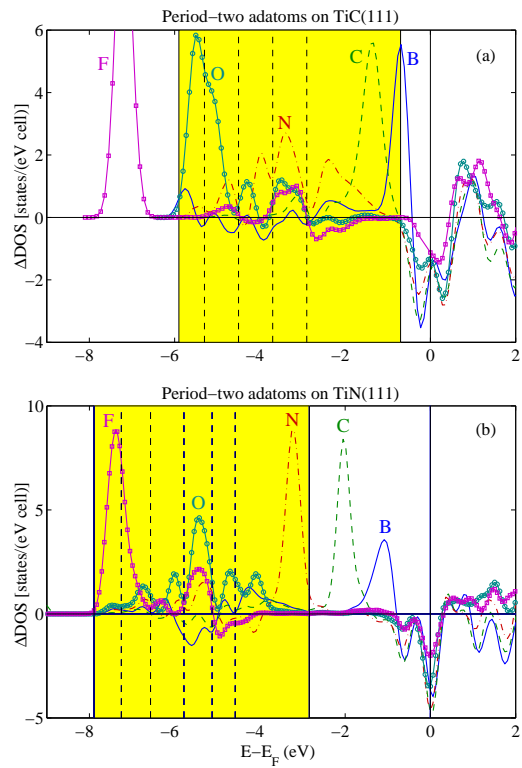}}
\caption{\label{fig:DeltaDOS}(Color online).  
Calculated $\Delta$DOS($E$) (difference in DOS for the surface TiX bilayer 
with and without adsorbate) for period-two adatoms in fcc site on TiX(111).  
Shaded regions mark the substrate UVB regions.  Dashed vertical lines mark 
the energies of the substrate UVB peaks.}
\end{figure}

{\it (i) Increasing the adatom group number within each period}
downshifts the adatom energy level.  The adlevel--TiSR energy matching is 
thus successively reduced.  Indeed, the negative $\Delta$DOS peaks 
at $E_F$ decrease successively as group IV$\rightarrow$VII 
(Fig.\ 3), indicating a weakening adlevel--TiSR coupling.  
On the other hand, the adlevel--XSR matching increases as group 
III$\rightarrow$VI.  According to our $\Delta$DOS's and 
DOS(${\bf r}, E$)'s, this strengthens successively the 
adatom--XSR couplings, as exemplified by period-two adatoms on 
TiC(111) [Fig.\ 3(a) and Ref.\ 
\onlinecite{Ruberto_Vojvodic_Lundqvist}]: 
for B, the effective adatom $p$ level lies at the upper UVB edge and is 
very sharp, indicating no coupling with UVB states; 
for C, the adatom $p$ state lies in the upper UVB region, is 
slightly broadened, and forms a small, weakly bonding, adatom--C subpeak 
just below CSR4, indicating a weak adatom--UVB coupling; 
for N, the situation is similar to H, the adlevel lying in the middle of 
the UVB and strongly broadening and splitting into subpeaks of mixed 
bonding and antibonding adatom--C character (more bonding at lower energies); 
for O, the adlevel lies in the CSR-dominated UVB region and forms strong 
adatom--C bonds.  
Similar results are obtained for period-2 adatoms on TiN(111) 
[Fig.\ 3(b)] and period-3 adatoms on both TiX(111) 
surfaces \cite{Ruberto_Vojvodic_Lundqvist}.  
Thus, the increasing group III$\rightarrow$VI $E_{\rm ads}$ trend
follows closely the increasing adlevel--XSR coupling trend, while very poor 
correlation is found with the adlevel--TiSR trend.  
The opposing adlevel--XSR and adlevel--TiSR trends can, however, 
explain the almost constant $E_{\rm ads}$ value as C$\rightarrow$N.  

{\it (ii) Increasing the adatom period number within each group}
increases the adatom radius and upshifts the adlevel, due to 
core orthogonalization.  The first effect reduces adlevel--SR 
wavefunction overlaps.  Our $\Delta$DOS's show that this 
affects the adatom--TiSR coupling strength, decreasing the 
negative peaks at $E_F$ \cite{Ruberto_Vojvodic_Lundqvist}.  
The second effect changes the adlevel--SR energy matchings.  
Our $\Delta$DOS's and DOS(${\bf r}, E$)'s 
show that this weakens the adatom--XSR coupling strengths, 
reducing the broadening and splitting of the TiSR-modified adlevel 
\cite{Ruberto_Vojvodic_Lundqvist}.  
Thus, the decreasing period 2$\rightarrow$3 $E_{\rm ads}$ trend follows 
closely these reductions in both adatom--TiSR and adatom--XSR coupling 
strengths.  

{\it (iii) Changing the substrate from TiC to TiN(111)} 
causes a downshift of all SR's and an increase of the clean-substrate TiSR 
filling.  For a given adatom, the first effect reduces the adlevel--XSR 
matching, thus weakening the adatom--XSR couplings 
(Fig.\ 3).  
The second effect strengthens the adatom--TiSR coupling, 
as evidenced by the larger depletion of TiSR electrons on TiN 
(Fig.\ 3).  
For low adatom-group numbers, the adlevel--XSR matching is 
negligible on both surfaces and the adatom--TiSR coupling dominates, thus 
favoring adsorption on TiN.  As the group number increases, the adatom--TiSR 
coupling weakens in favor of the adatom--XSR couplings, causing 
successively stronger adsorption on TiC.  Hence, the crossing 
of the TiC and TiN $E_{\rm ads}$ trends is well explained by the opposing 
trends in adlevel--TiSR and adlevel--XSR matchings as the group 
number increases.  

Thus, the $E_{\rm ads}$ trends are well correlated with electron-structure 
trends in which both types of SR's (TiSR and XSR's) are considered.  This 
is further supported by: 

{\it (iv) The weaker chemisorption at the end of each adatom period} 
does not follow the energy-matching trends of (i) above.  
Yet, sharper $\Delta$DOS $p$ peaks for F and Cl, compared to 
other adatoms with same adlevel--XSR energy matchings 
(Fig.\ 3 and Ref.\ \onlinecite{Ruberto_Vojvodic_Lundqvist}), 
indicate weaker adatom--XSR couplings for group-VII adatoms.  
The electronic structure shows a change in adatom--TiSR bond nature 
from iono-covalent to ionic as group VI$\rightarrow$VII: 
smaller TiSR quenchings (Fig.\ 3), 
almost spherical adatom valence-electron distributions (Fig.\ 4), 
Bader charge transfers to the adatoms of 0.75--0.80 electrons, mostly from 
the nearest-neighbor Ti atoms, 
and mainly adatom-centered DOS(${\bf r}, E$)'s 
at the adatom-peak energies for F and Cl.  Thus, the weaker adatom--XSR 
couplings for F and Cl are caused by their almost completely filled valence 
shells, which are essentially inert to interaction with the UVB.  Also, 
for F/TiC(111) the adlevel lies well below the substrate UVB, yielding 
further adatom--XSR weakening.  
These effects weaken the chemisorption as group VI$\rightarrow$VII.

\begin{figure}
\scalebox{.38}{\includegraphics{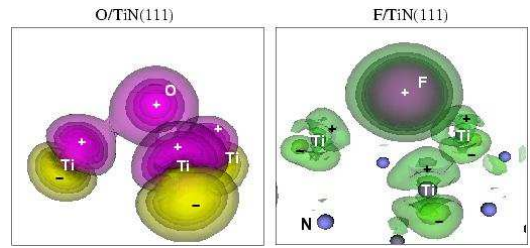}}
\caption{\label{fig:CHD}(Color online).  
Isosurfaces of adsorption-induced differences in electronic charge 
densities (magenta, $+$ = increase; yellow, $-$ = decrease; 
green = no change) for fcc O and F on TiN(111).  The right panel shows a 
larger number of isosurfaces, illustrating also smaller absolute values of 
the charge-density difference, than the left panel.}
\end{figure}

{\it (v) The much weaker chemisorption on TiC(001)} 
evidences the importance of the adatom--TiSR coupling, as 
TiC(001) lacks TiSR at $E_F$ 
\cite{Zaima,Oshima001,Ruberto_Vojvodic_Lundqvist}.  

{\it (vi) The stronger adsorption in fcc than hcp}
follows from the TiSR spatial extent [Fig.\ 2(a)], 
which favors adlevel--TiSR 
wavefunction overlap and thus adatom--TiSR coupling in fcc.  
In turn, this causes a larger downshift of the TiSR-modified adlevel 
compared to hcp, affecting the adlevel--XSR energy matching and thus also 
the adatom--XSR coupling strengths (as shown for O/TiC in 
Ref.\ \onlinecite{Ruberto_Vojvodic_Lundqvist}).  
The largest such effect should be 
obtained for adatoms with adlevel close to the XSR's.  
Indeed, our calculated fcc--hcp $E_{\rm ads}$ differences show pyramidic 
trends that are very similar to those for the fcc $E_{\rm ads}$ values 
(with peaks, for TiC, at N and S \cite{Ruberto_Vojvodic_Lundqvist}), 
showing again a very good correlation between calculated $E_{\rm ads}$ 
values and adlevel--XSR matching trends.  It can be expected that 
diffusion-barrier variations (Fig.\ 1) are explainable 
in a similar way.  

Thus, the CCM well describes the key features of the $E_{\rm ads}$ 
trends on TiX.  The large adsorptive variety is caused by the concerted 
action of adatom couplings to both surface TiSR and subsurface XSR's. 
The model should apply also to 
other systems with similar electronic structures, such as 
other TMX's, and thus have broad technological implications.  

Financial support from VR and SFS via ATOMICS and the allocation of 
computer resources at UNICC and SNIC are gratefully acknowledged.

\vspace*{-1em}



\end{document}